# A Local Realist Account of the Weihs *et al* EPRB Experiment

Donald A. Graft

donald.graft@cantab.net

Quantum mechanics stands in conflict with local realism only in its treatment of separated systems. A modification of quantum mechanics that changes the handling of separated systems is suggested that can reconcile quantum mechanics with local realism. An apparent obstacle to this program is the experimental evidence, but I argue that the experiments have been misinterpreted. By way of example, I describe a local realistic account of one important EPRB experiment that is claimed to demonstrate nonlocal entanglement. The local model can be calibrated into both quantum and classical domains via adjustment of parameters ('hidden variables') of the apparatus. Weihs incorrectly dismisses these parameters as uncritical, whereas we show that device calibration is crucial. When properly interpreted, the experiments show that nonlocal entanglement is an error. The rest of quantum mechanics remains intact, and remains highly valued as a powerful probability calculus for observables. Quantum mechanics and local realism can be reconciled, and they each can offer useful paradigms for describing systems.

## INTRODUCTION

The apparent nonlocality of quantum correlation (entanglement) has confounded all attempts to reconcile it with other basic laws of physics, such as Lorentz invariance. It is not hard to see why this should be so. The standard quantum mechanics (QM) prediction for the probability of measuring both photons as 'up', for separated measurements of the correlated singlet state at stations A and B, is believed to be given by a joint distribution, which I call $QM(AB)_J$, where J stands for joint and $\theta$ is the angle between the analyzer settings at A and B, $\alpha$ and $\beta$, respectively :

$$QM(AB)_J = P(AB \mid \alpha, \beta) = \tfrac{1}{2}\cos^2(\theta)$$

The A and B events are correlated as determined by the singlet state. The measurements that are taken in an EPRB experiment, however, are separated, so that the measurement at A proceeds in ignorance of the analyzer setting at B and the outcomes at B, and vice versa. Therefore, we expect to measure the marginal probabilities at A and B, so that the prediction is:

$$QM(AB)_M = P(A \mid \alpha)P(B \mid \beta)$$

$QM(AB)_M$ represents the predicted probability using the marginals $P(A \mid \alpha)$ and $P(B \mid \beta)$. Strangely, it is generally believed that the prediction $QM(AB)_J$ can be measured via separated sampling of A and B. But this leads to a contradiction, because it means that:

$$QM(AB)_J = QM(AB)_M$$

or:

$$P(AB \mid \alpha, \beta) = P(A \mid \alpha)P(B \mid \beta)$$

That is the condition for independence of A and B, but we know their dependence (correlation) was assured by the singlet state. This constitutes a *reductio ad absurdum* of the idea that a joint probability can be sampled by means of marginal (separated) measurements. Similar considerations apply for the anticorrelated singlet state.

It is possible to escape this *reductio* by supposing that when a photon is first measured at A, the companion photon at B is 'projected' to a state that will yield a measurement compatible with $QM(AB)_J$, that is, that the prediction is in the form $P(A|\alpha)P'(B|\beta)$, where $P'(B|\beta)$ represents statistics from the measurement of projected photons. The *reductio* is dissolved, but at a terrible price: we have to accept superluminal effects of a strange kind for which no physical mechanism is known. We also have to ask which photon projects the other. Is it the first one to be measured? That is impossible to decide when events have different orders in different reference frames. From an engineering perspective, it is challenging to conceive a protocol that ensures that one and only one of the photons projects the other, and the resulting mechanism has to be regarded as fanciful.

The direct way to resolve this terrible dilemma of modern physics appears to be to accept that a joint distribution cannot be sampled by means of marginal (separated) measurements. One must use $QM(AB)_M$ instead of $QM(AB)_J$ for predicting the measured correlation. This prediction can be made via reduced density matrices in a manner completely analogous to that of marginalization and conditional probabilities in standard probability theory. If this is accepted, then a very small modification of QM, that is, using the marginals versus the joint probability in separated measurement situations (exactly as in classical probability), can reconcile QM and local realism. I have more to say about this reconciliation later in this paper.

There are flies in the ointment for advancing this program – the experiments! Modern consensus is that the results of EPRB and other experiments confirm that the $QM(AB)_J$ prediction is clearly obtained. Nevertheless, I argue that the experiments have been misinterpreted. I analyze a careful and oft-cited EPRB experiment by Weihs *et al*[1], hereafter referred to simply as Weihs, and show that there exists a *plausible* local realistic account of the experiment's results. The results are shown to approximate the $QM(AB)_J$ solution just as the Weihs experiment does, but as a result of local realistic mechanisms rather than the dubious sampling of a joint probability via its marginals implied by the $QM(AB)_J$ prediction. While attempting to deconstruct all of the experiments can quickly turn into an exasperating game of *whack-a-mole*, we will see that the mechanism in effect in the Weihs experiment has broad applicability, and so a plausible local realistic account of the Weihs experiment goes a long way toward clearing the path for reconciliation between QM and local reality. We turn first, therefore, to a description of the local realistic account of the Weihs experiment.

## LOCAL REALISTIC ACCOUNT OF THE WEIHS EPRB EXPERIMENT

### *Testing the QM Prediction*

The QM prediction $QM(AB)_J$ implies several consequences that can be tested experimentally and which therefore can be regarded as criteria to be satisfied by any local realistic model that seeks to account for the results of the Weihs experiment:

1. Results close to the predicted joint and marginal probabilities should be observed.

2. The observed probabilities, coincidence counts, and singles counts should be rotationally invariant.

3. Criteria 1 and 2 should be observable with complete symmetry of the two sides of the experiment.

It is still widely believed that the Bell and related inequalities have proven that no local realistic model can satisfy the first criterion, i.e., that no local realistic model for separated measurements can produce the $QM(AB)_J$ joint and marginal probabilities. However, several models[2,3,4,5] based on variable detection have generated QM statistics violating Bell inequalities and so have decisively shown that the first criterion alone is not sufficient to experimentally prove the validity of the $QM(AB)_J$ prediction. An additional criterion is needed to distinguish QM from local realism.

It is also widely believed that all local realistic models depending on variable detection must exhibit rotational variation of the observed statistics (note that I use the term 'rotational invariance' in a different and wider sense than is

                                                                 

usual; this seems more convenient than introducing a new term). For example, variation of the total number of detected coincidences is observed as the measurement angle difference is scanned through the range 0 to 180 degrees, whereas quantum mechanics predicts that the number of detected coincidences should be constant. Again however, local realistic models[6,7] exist that can generate the quantum statistics with full rotational invariance, and therefore a further criterion is needed.

So we arrive at the final redoubt of the no-go arguments: symmetry. It is widely believed that any local realistic model satisfying the first two criteria must be grossly asymmetric in a manner that is so implausible as to disqualify it from serious consideration. I show here on the contrary that any real experiment includes broken symmetries in some factors affecting the measurements and that it is very easy to attain quantum-like behavior from a local realistic model operating on a broken symmetry in a very natural and plausible way. The perspective being developed here goes beyond providing a local realistic account of the Weihs experiment; it suggests a way to reconcile the QM prediction and local reality. They are seen to operate in different domains of the parameter space of the experiment. In the account of the Weihs experiment to be presented, a single parameter will be seen to continuously vary the system behavior between quantum-like and local realistic behavior, and a plausible experimental calibration can place the system in a quantum domain.

## *Description of the Model: Light Source*

To represent the light signal (photon) pairs produced in the Weihs experiment by a pumped parametric down conversion (PDC) crystal and associated optics, we choose the following classical mechanism:

```
forever
  // Generate 'entangled' photon pair.
  theta1 = (2 * PI * ran1())
  theta2 = theta1 + PI/2 + ran2(decoherence)
  if (theta2 > 2 * PI)
    theta2 -= 2 * PI
```

This mechanism emits a stream of paired anticorrelated pulses (orthogonally polarized at angles *theta1* and *theta2*) uniformly distributed around the axis of propagation. The strict anticorrelation (90-degree separation between *theta1* and *theta2*) can be relaxed by specifying a *decoherence* factor. This is a fully rotationally isotropic source. In a following paper I consider light sources constrained to a fixed polarization base, such as H/V, and other reduced rotational symmetries.

## *Description of the Model: Detector*

To represent the detection of light pulses (photons) in the Weihs experiment by means of electro-optic modulators (EOMs), Wollaston prisms, avalanche photodiodes (APDs), and constant-fraction discriminators (CFDs), we choose the following classical mechanism based on energy splitting in a polarizing beamsplitter:

```
up = down = 0
if (ran1() < efficiency)
  f = cos(A - theta1) * cos(A - theta1) +
      noise_amplitude * ran2(noise_variance)
  if (f > thresh)
    up = 1
  f = 1 - f
  if (f > thresh)
    down = 1
```

This mechanism implements an inefficient classical polarization beamsplitter oriented at angle A and following Malus's Law. The *efficiency* parameter allows for loss of source events due perhaps to detector dead times, losses in the optical pathways, etc. Although the account described here does not rely on this inefficiency, it is included for two reasons. First, real experiments typically have low efficiencies. Second, it avoids a possible objection to the scheme to be described, i.e., that it is implausible to suppose that a detector detects all events. We will see that the effect to be described requires only that one of the sides detects a *fair sample* of the source events, and not that it detects *all* source events. We return to this point later as it is relevant to understanding when rotational invariance can be expected. Note that all the simulations presented use an arbitrary efficiency of 0.5 (50%). The account does not rely on this or any other specific value, however.

Dual-channel detectors (up and down channels) each compare against a defined threshold to detect a light pulse. The mechanism allows for a noise contribution prior to the beamsplitting, representing experimental conditions. Here the *thresh* parameter directly corresponds to the detector CFD threshold in the Weihs experiment. Its calibration is important and must be carefully considered.

Note that we do not explicitly represent the electro-optic polarization modulators (EOMs) in the Weihs experiment. An EOM is used to set the effective measurement angle by rotating the polarization of the source photons while keeping the Wollaston prism beamsplitter and APD detectors in a fixed position. Modulation of the rotation can be performed very rapidly using the EOMs, compared to physically rotating the prism and associated detectors. For our purposes, we stipulate that the EOMs perform a rotation equivalent to physically rotating the prism and detectors, as intended by Weihs. It is conceivable that the EOMs impose some other unintended effects, such as amplitude or phase modulation, conversion of linear polarization to elliptical polarization, etc. But we do not require such effects to account for the results of the Weihs experiment, and so we assume that the EOMs are performing as intended and we do not explicitly model them.

We also do not include the paradigm of time-tagged data collection and subsequent temporal window filtering used by Weihs. My analyses of the data do not show any unfair sampling due to coincidence-window modulation, in agreement with previous published analyses of the data[8,9]. So we choose to stipulate that the windowing mechanism in the actual experiment delivers a fair sample of source events and we choose not to explicitly model it.

We assign the same threshold to both the up and down detectors at each side to simplify matters and minimize the number of parameters of the experiment. In a real experiment, there would be four thresholds to calibrate. We require only the simplified model to account for the Weihs experiment, so we demonstrate that and reserve the option to later consider the effects of a symmetry breaking between the up and down channel thresholds. A single symmetry breaking between the two sides is enough to account for the Weihs data.

Finally, the noise model must be considered. The model to be presented will rely upon one of the sides being fairly sampled by the measurements and so we must be sure that noise is not leading to unfair sampling (at least at one side). Noise generally reduces the visibility of the correlation, but it is also possible for noise to produce unfair sampling (bias), through the mechanism of stochastic resonance. These effects are not required to demonstrate a local-realist account of the Weihs experiment, so we include only some noise needed to produce physically realistic correlation curves. The Weihs experiment succeeds in keeping noise low, both with a large signal-to-noise ratio at the detectors, and with an effective time-windowing filter applied to the raw data during data analysis. The source decoherence is not known as Weihs did not perform tomography of the emitted source pairs.

The model here generates individual detection events from an objective and deterministic mechanism, but the measurements remain stochastic due to inclusion of randomness. The source of randomness in the model and, it is argued, in the experiments, is uncontrolled variability in the source coherence, variability in source light pulse amplitudes (or energies), and background electromagnetic noise at relevant frequencies throughout the experiment.

## *Description of the Model: Calibrating the Detector*

We can connect our light source to our detector and observe the results as a function of the threshold. For each source emission, we will have one of the following results: both the up and down channels register a count ('double'), neither the up nor down channels register a count ('miss'), the up channel alone registers a count ('up'), or the down channel registers a count ('down'). In a real experiment we would not know about the misses, because we would not know the number of source emissions. Figure 1 shows a plot of the results with *efficiency* = 0.5, *noise_amplitude* = 0.3, and *noise_variance* = 0.2. The measurement angle A is set to 0 degrees, but the rotational invariance of the source ensures that similar results are obtained for any meaurement angle. We render the misses in a dashed line to remind us that we wouldn't know the number of misses in a real experiment. At a threshold of 0.5, the number of doubles is reduced to zero and the number of up/down events is maximized. The correct threshold to set for our detector is evidently 0.5, in agreement with theoretical considerations.

The reduction of doubles to 0 at threshold

0.5 is unphysical but inconsequential. Inclusion of dark counts would correct this by producing a low background of doubles, and eliminate this artifact of our model, but this is not necessary to demonstrate our local-realist model. Later we bring back dark counts for another purpose.

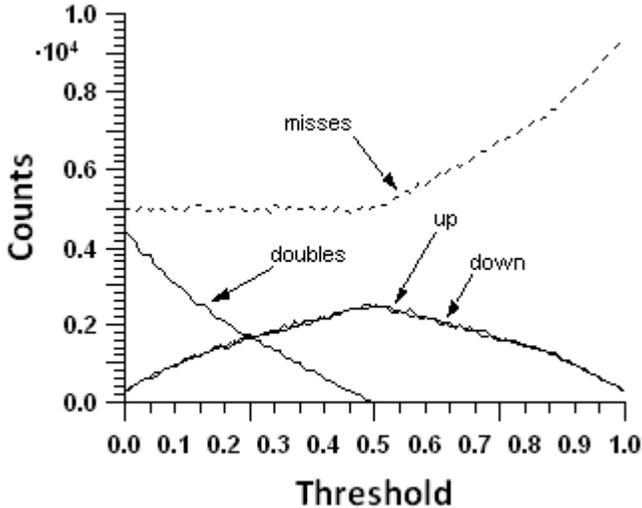

*Figure 1.* Detector calibration plots

It is important to realize that for simplicity we are dealing here with light pulse amplitudes normalized to 1.0, and that makes 0.5 special as it is half the signal amplitude. In a real experiment the light pulses have unknown amplitudes, and so the calibration criterion becomes setting the threshold to half the signal amplitude. As the latter is not known, setting the threshold becomes a challenging matter. We will return to this when we consider how Weihs may have calibrated his experiment. Meanwhile, we continue in normalized amplitudes.

## *Description of the Model: Correlation of Detectors*

We can connect two separated detectors to our source in the usual EPRB fashion and observe the correlation of the results from the two detectors. We discard events where either detector reports a double (up and down channels both report a pulse). We also discard events where either detector reports a miss (neither the up nor the down channel report a pulse). This is a natural post-selection that appears reasonable and is consistent with what is known about the details of the Weihs experiment. The remaining events are 'coincidences' where both detectors have each reported an up or down detection.

When the two sides report opposite results (up versus down) we register a match, because we expected anticorrelation. When they agree we register a mismatch. Then our correlation metric will be given by:

$$P = matches/(matches + mismatches)$$

QM predicts P will vary as $\cos^2\theta$, where $\theta$ is the difference between the angle settings of the two detectors. (The metric P should not be confused with the expectation E, which varies from -1 to 1 and involves $\cos 2\theta$ rather than $\cos^2\theta$. There is a straightforward mapping between expectation E and match probability P; they are equivalent formulations. The probability metric P is preferred here to more easily connect it to probability theory and to our argument in the Introduction.)

The parameters of our model, other than the two angles of measurement determining $\theta$, consist of efficiency, noise, and decoherence, and importantly, the CFD threshold parameters at both sides of the experiment. We choose some plausible values for efficiency (0.5), noise (amplitude 0.3, variance 0.2), and decoherence (0.2). As experimenters we are now faced with how to set the free threshold parameters. A simple approach is to calibrate the thresholds while observing the correlation with the aim of finding quantum correlations. So we start with the thresholds randomly set and then start tweaking them as we look at the resulting correlation curves. To do this in our normalized local realistic model, we select (for example) thresholds 0.3 at A and 0.7 at B. The correlation results are displayed in Figure 2. The measurement angle is held at 0 degrees at side A and the angle at side B is varied from 0 to 180 degrees. The match probability P is seen to be promising as we seek a $\cos^2\theta$ curve.

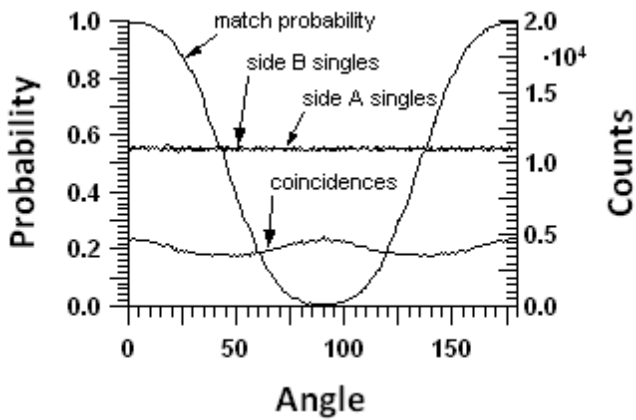

***Figure 2.** Correlation resulting from arbitrary, uncalibrated thresholds*

Looking at the coincidences curve of Figure 2, we note a marked rotational variation of total coincidences in contradiction to quantum mechanics. Our correct CFD thresholds are not known, so we start to calibrate by changing the threshold on side A as we look at the resulting curves. We rapidly find that at side A, threshold value 0.5, we keep the promising correlation curve while we eliminate all rotational variance. So we choose to set threshold A to 0.5. The resulting threshold corresponds to the correct calibration of a single detector already described, so we could have arrived here by a formal calibration following that procedure. The resulting correlation can be seen in Figure 3.

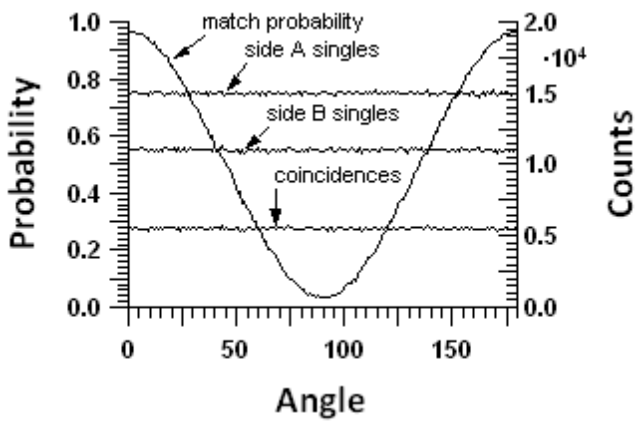

***Figure 3.** Correlation resulting from correct calibration of one side*

Now we need to calibrate side B by adjusting its threshold as we look at the resulting curves. We notice that the visibility of the correlation curve in Figure 3 is a bit low (QM predicts 100% and 98% is seen in the Weihs experiment). We quickly find that raising the threshold at side B to 0.75 (or lowering it to 0.25) improves the visibility of the correlation curve and allows us to collect the high-visibility, rotationally invariant quantum statistics shown in the top of Figure 4 that we seek. To fully show the rotational invariance, the bottom of Figure 4 shows the same settings but with side A's angle offset by 45 degrees.

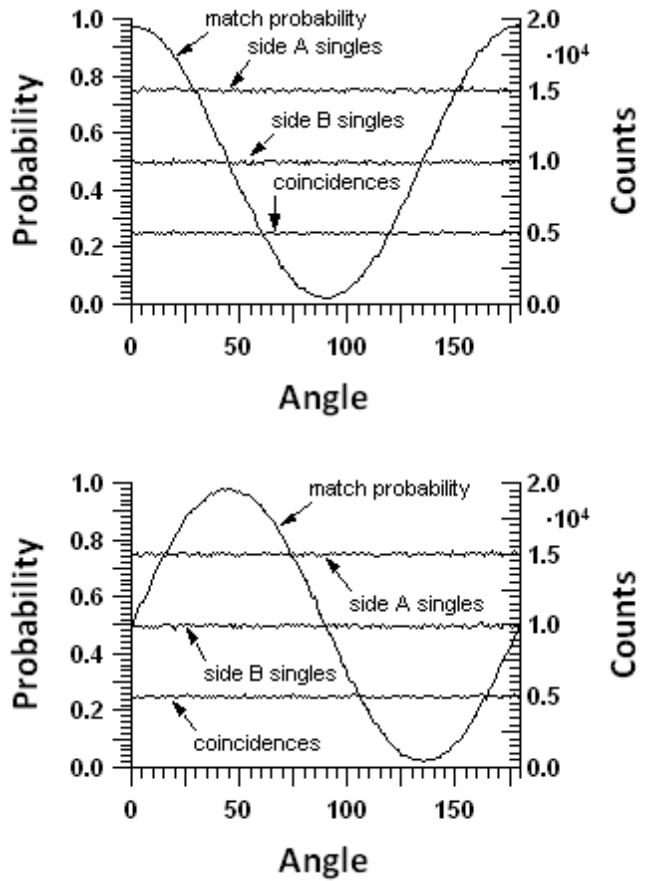

***Figure 4.** Quantum correlation after sequential calibration of the sides (top: A = 0 degrees, bottom: A = 45 degrees)*

A different, more rigorous calibration of the experiment also leads to the same result. We calibrate both sides separately so that we start with threshold 0.5 at both sides. The resulting curves are shown in Figure 5. The visibility is low and Bell's Inequality is satisfied. To morph the system toward quantum correlation we can choose to simply adjust either side's threshold in either direction! Any one such adjustment can continuously vary the correlation statistics from classical (Figure 5) to quantum (Figure 4). We could set 0.75 at one side as before with very good results, but 0.25 also produces similar results. If we push the adjustment too far we reach super-quantum correlations, shown in Figure 6 (0.5/0.92).

Of course it is interesting to know from Weihs himself how the thresholds were calibrated. Weihs writes as follows about the CFD threshold setting[10]: "The choice of this threshold is for avalanche photodiodes very uncritical, because the detection pulses have much higher amplitudes than the thermal noise." The published material does not further address this matter and there is no reporting or justification of the threshold settings for all four detectors actually used in the experiment, nor any study of the effects of different thresholds on the results. In a recent email exchange[11] Weihs doesn't recall exactly how the thresholds were set, and repeats that one needs only to set the threshold to a level above the noise. That leaves open as possibilities all the calibrations discussed here. In the next section we argue that Weihs may have calibrated one side's threshold too high, close to the threshold value 0.75 as defined here. The mechanism of unfair sampling acting in the experiment will be seen to rely on misses and not doubles.

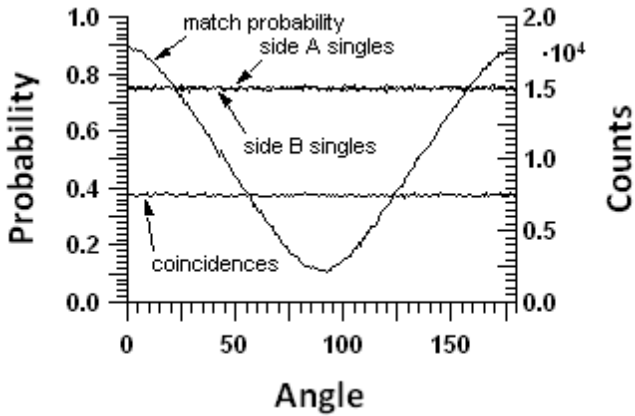

*Figure 5.* Classical correlation after independent calibration of the sides

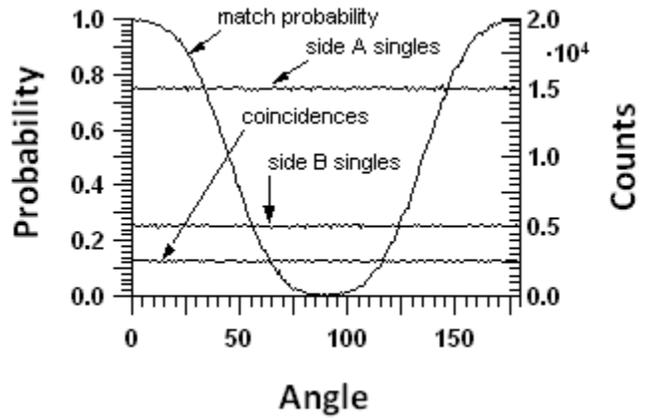

*Figure 6.* Super-quantum correlation

## Signatures of the Local Model in the Weihs Data

A prominent feature of our model in its quantum calibration is the difference in singles counts between the two sides (Figure 4). This results from the asymmetry of threshold settings between the sides. We would expect to see this feature in the Weihs data if it were produced by the mechanism of our model. Indeed we see this artifact in the Weihs data, as seen in Figure 7, reproduced from Adenier and Khrennikov[12]. We see the asymmetry between the sides overall as we expect, but we also see a further symmetry breaking between the up and down detectors at Alice (which we could model with separate thresholds at the up and down detectors). This asymmetry of singles counts between the sides is not predicted by quantum mechanics and must be regarded as evidence for the operation of our local-realist mechanism.

        

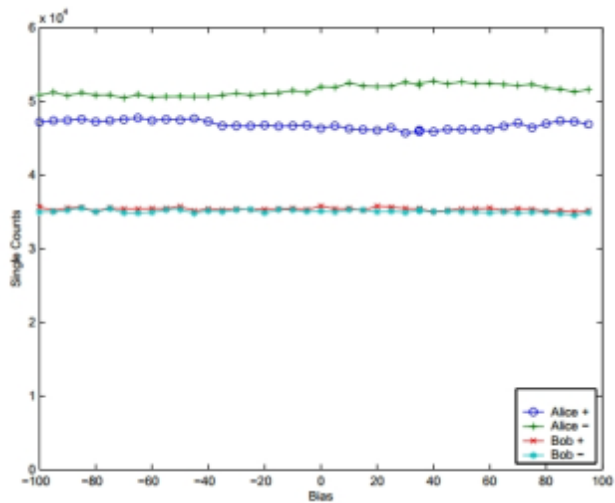

***Figure 7.*** *Singles in the Weihs data as analyzed by Adenier and Khrennikov[12]*

Furthermore, this large asymmetry in the Weihs experiment violates our 3rd criterion for testing the QM prediction (that the statistics are observable with fully symmetric sides). In fact, the experiment succeeds only as a didactic demonstration of how easily variable detection (unfair sampling) can sneak into an experiment and totally mislead us. Any significant asymmetry in the reported data of an experiment must be treated with extreme caution, and ruthlessly eliminated if a decisive conclusion is desired. We have seen that symmetry is the decisive criterion and so any experiment that seeks to establish the QM prediction must be manifestly symmetric in all relevant observables of the experiment.

Weihs initially suggested[11] that nonmaximal entanglement may account for the observed large counts asymmetry, arguing that with nonmaximal sources one may observe "more H than V". We can accept that nonmaximal states may produce a difference between H and V counts at a side, but that difference should be the same magnitude at both sides (because they share common source singlets). However, in the *scanblue* data, one side appears maximal (H and V counts are close to each other) while the other side appears markedly nonmaximal (H and V counts differ significantly). Worse, this effect could not produce the observed large asymmetry of total singles (H plus V) between the two sides. Later[11], Weihs accepts that, but still looking for an explanation for the large difference in singles counts between the sides, suggests that it arises from large differences in the efficiency of the fiber couplings, and other factors. If the experiment indeed contains such large uncontrolled factors producing asymmetry between the sides, it can hardly be definitive in distinguishing the (standard) quantum prediction from the local realistic one. The true state of affairs can be tested by varying the threshold at one side and observing the effect on the counts asymmetry, but Weihs did not report on the effects of threshold variation.

My analysis of the Weihs data reveals a further asymmetry in double detections between the sides that can be accounted for by our model. A double detection is a simultaneous up and down detection at one side. We consider doubles to be detections on the same side that are very close to each other in time and which are opposite (one up, one down). With a small enough window, the detector dead time precludes seeing any doubles with both up or both down. In totaling all runs of the Weihs *scanblue* data, with a window size of 4ns, I found 379 doubles at side A and 181 at side B. This asymmetry is not predicted by quantum mechanics. We realize however that doubles can be produced by fortuitous dark events and that the rate of dark events depends on the detector threshold. The threshold asymmetry of our model modulates dark events in a manner that accounts for the observed asymmetry in double detections.

Importantly, we also notice that the absolute number of doubles is very low and therefore insufficient to produce significant unfair sampling. We are thus led to believe that Weihs inadvertently calibrated his threshold too high on one side, rather than too low. The unfair sampling is due to misses rather than doubles, and the few doubles we see are due to dark counts. Misses can obviously be seen only by their effect on the singles rates.

While speaking of doubles I report that in the Weihs experiment the effective dead time of the detectors can be assessed by looking for closely spaced detections in one detector as a function of the coincidence window size. Below about 0.7μsec, there are no such pairs, suggesting that the dead time is close to the 1 μsec reported by Weihs.

We have discussed several significant deviations from the predictions of (standard) quantum mechanics in the data from the Weihs experiment. De Raedt *et al*[13] more formally address the likelihood that the quantum prediction can account for the data of the Weihs experiment. They conclude that it is unlikely.

## DISCUSSION

This paper has presented a local realistic system (experiment) that can be calibrated into both quantum and classical domains via adjustment of parameters ('hidden variables') of the apparatus. By incorrectly dismissing these parameters as uncritical, and by omitting to report the values of these parameters and their effects on the correlation curve, Weihs missed the opportunity to decisively confirm local reality. Indeed, experimenters can at will use an EPRB experiment to either confirm classical correlations or quantum correlations, by innocently varying experimental parameters that are thought to be uncritical, as part of a reasonable calibration procedure for the experiment. Interestingly, the results of Holt and Pipkin[14], that disconfirmed the QM prediction for correlations, were never published, apparently due to scientific peer pressure. No convincing invalidation of the Holt and Pipkin results has been published. We argue that Holt and Pipkin used an independent calibration procedure for the sides that invoked the classical domain (calibration) of our model. The fashion nowadays, of course, is to "verify nonlocality" by reporting only the results of experiments calibrated to show quantum-like correlation.

Weihs too apparently succumbs to this mindset of seeking and reporting only quantum results. While he does not address calibration of the CFD thresholds, which he considers to be uncritical, he does discuss alignment of the compensator crystals[15]: "By tilting the compensator crystal in the source we were able to optimize the correlations to those of the singlet state." Weihs appeared willing to calibrate his experiment while observing correlations rather than by individually calibrating each side correctly and independently prior to gathering correlation data. When this approach is applied to the CFD thresholds, combined with the criterion of success being demonstration of optimal quantum-like correlation, the result is inevitably to bias the thresholds into a domain that shows quantum-like correlations, as we have shown.

The mechanism described here has broad applicability to many other EPRB experiments. Specifically, any detectors having thresholds, or any other physical asymmetries that affect detection, are immediately suspected. Light detectors in the photon experiments typically use a photomultiplier tube (PMT), avalanche photodiode (APD), superconducting transition edge sensor (TES), or other mechanism to capture and amplify the incident electromagnetic energy and represent it in analog form, followed by a threshold-discriminator of various designs to produce detection events. The highly efficient TES detectors (98% energy capture), that can reliably distinguish single photons, still must be thresholded to define the boundary lines between photons (light pulses) in the analog trace of the detection, and to distinguish a single photon from the noise. The efficiency of the energy capture is irrelevant if the thresholding mechanism (or other relevant parameter of the apparatus) subsequently imposes unfair sampling. Our account is directly applicable to these photon experiments. Atom experiments may also use thresholded detectors that can exhibit the behavior of our mechanism. Responding to the suggestion that the mechanism described here may apply widely to EPRB experiments, Weihs effectively concedes the point[11]: "I agree there will always be some kind of threshold or discrimination..."

A recent experiment[16] attempts to completely bypass all unfair sampling mechanisms by invoking Eberhard's inequality, which explicitly includes missed detections[17]. However, such experiments don't necessarily avoid the mechanism of this paper. In a semiclassical model with low thresholds, we may obtain detections in both channels at a given side. Eberhard's inequality includes *o* (ordinary path), *e* (extraordinary path), and *u* (undetected) events at a given side, but omits *oe* double detections at a side. There may also be a need to distinguish *o* from *ou* and *e* from *eu* events at a side. Discarding or misclassifying these events generates unfair sampling, as shown here. This paper is not the place to deconstruct the experiment[16] but I remark that it is easy to demonstrate a local realist account based on these ideas.

It is important to realize that we don't need inequalities like Eberhard's to test the (standard) quantum prediction for coincidences. Detection inefficiency need not necessarily lead to unfair sampling. Even an experiment as inefficient as the Weihs experiment (efficiency = 0.05) could suffice, as long as the two sides are properly calibrated to deliver a fair sample of the source events. I argue that if that had been done, the experiment would have confirmed local realism. Therefore, we should not minimize or neglect the important contribution of the Weihs experiment, despite its being incomplete. It is a careful and pioneering experiment that has the potential to discriminate between the (standard) quantum prediction and the local realist prediction, as long as symmetry is maintained in all important apparatus parameters through careful and correct apparatus calibration.

Quantum physicists may be uncomfortable with the model described here because they believe that 'photons' are indivisible particles that cannot be divided, for example at a beamsplitter. But the question of whether the electromagnetic field itself is quantized is still not settled. The current consensus that photons do not split at a beamsplitter is based on results such as those of Grangier *et. al.*[18], but the interpretation of the results relies on only one of many possible classical models of detection. Khrennikov[19] shows that a detector model using thresholds like the one described here easily accounts for photon anticorrelation at a beamsplitter. Indeed, we can easily see from Figure 1 that doubles drop to 0 at a threshold of 0.5, and because a double in our terminology is the same as a coincidence in the context of the Grangier experiment, semiclassical theory predicts anticorrelation for properly calibrated detectors. Khrennikov observes that Grangier admits to using a "rather high threshold", so we can be fairly sure that it was greater than or equal to 0.5, which is in a domain where the semiclassical model predicts total anticorrelation (neglecting dark counts). Grangier makes the same mistake as Weihs by dismissing detector thresholds as uncritical and failing to report the effect of modifying the thresholds on the observed statistics.

Adenier has proposed an interesting experiment to further address this issue, using analysis of analog TES traces[20].

In addition to the work just described, the role of detector thresholds in the EPRB context has previously been considered in important work of Hofer[21] and Adenier[22]. Neither apparently realized the importance of asymmetric thresholds and the role of fair sampling in producing rotational invariance. Adenier recently advocated searching for rotational variance in new experiments. That is a reasonable thing to do if one remembers that appropriate calibration is needed to show rotational variance, and that rotational variance can easily be calibrated away. Any new EPRB experiment should investigate and document the effects of apparatus calibration on the results of the experiment. If the calibrated parameters are indeed uncritical, the findings will bear that out. If not, we learn important things about our system. I showed here that if at least one side, due to its calibration, delivers fair sampling (and not necessarily perfect detection), then the experiment delivers full rotational invariance. The fair-sampled side marginalizes the other side.

Khrennikov has drawn attention to the necessity for correct threshold calibration[23] in experiments, and that is clearly an important focus, as we have seen. However, Khrennikov views the thresholds as important only for "discarding the contribution of the random background field" as part of a classical random theory, rather than as a straightforward mechanism of unfair sampling in a deterministic model, as developed here. Though not important for my argument here, I remark further as follows. I don't believe that anyone has proven that Nature must be either nonlocal or nonobjective. The model here is local and objective. I also do not agree that contextuality implies nonobjectivity. In any case these things are metaphysical considerations. Khrennikov advises to calibrate the threshold above the noise, just like Weihs. That is correct but not sufficiently precise. It must be calibrated to 0.5 times the signal amplitude and the noise must be small compared to the signal. It is not enough to remove the background; one has to ensure fair sampling. It is not true that use of thresholded detectors "ruins objectivity". I do not agree that the background can only be detected through joint measurement of correlated signals (there are straightforward classical ways to measure noise). Khrennikov presents no correlation/counts curves or simulations, so one may wonder what his model really does. There is no discussion of rotational invariance. Is the model fully invariant? We have seen that one needs invariance and symmetry to prove the standard quantum prediction. Khrennikov relies upon an interaction of low-level background (noise) with the signal. Presumably rejection of the noise would reject some signal as well, leading to unfair sampling. But there is not enough noise in the Weihs experiment to reproduce the quantum curves with this mechanism. The model may be sufficient to violate an inequality, but if it doesn't have a good global match to the quantum curves with full rotational invariance, then it is arguably not a viable candidate for serious consideration.

Earlier published models using asymmetry of detectors have been rejected as ugly and unphysical. For example, Maudlin criticizes his own 'Scheme B' model on the grounds that it shows rotational variance and that there is no plausible physical counterpart to the detection patterns. Then he writes[24]: "For all these reasons simulator schemes are unnatural and ugly." Yet we have shown that asymmetry need not be unphysical, that variable detection can arise from known physical mechanisms (Malus's Law), and that rotational invariance can be easily achieved through calibration of the experiment. Broken symmetries in physical factors affecting the measurements can bias the system into different domains of behavior, as we have seen with the CFD threshold. The model developed here is one that can and must be taken seriously.

From the hardcore realist perspective, the model here is classical realist all the way through, and the quantum-like behavior is just the result of a classical realist system operating in a particular domain of the apparatus parameters. On the other hand, it does give meaning to the concept of quantum correlations; they are the ones we get from a particular calibration. Or, from a Bayesian point of view, our prior for calibration might reasonably be one that gives a quantum result halfway between the two extremes of classical and super-quantum behavior, so in a sense one can say the expected result for an EPRB experiment is quantum correlation. But that is candy coating.

## CONCLUSION

The contributions of this paper are:

1. A plausible local account of the Weihs experiment is presented and demonstrated with a computer simulation.

2. It is demonstrated that rotational variation is not a necessary outcome of unfair sampling. The strength of the Weihs experiment was its clear rotational invariance, and no plausible model had duplicated that. The model described here does so and identifies a 'smoking gun' in the experiment.

3. The critical importance of device calibration and reporting of all apparatus parameters is shown, and recommendations are made for experimental design.

4. It is shown that Eberhard's inequality fails to fully eliminate unfair sampling because it corrects for misses but not for doubles. Experiments relying on it are therefore inconclusive.

5. It is argued that high detection efficiency is not needed to distinguish quantum mechanics and local realism. This is significant in my view because the debate has thus far been distracted by a misguided focus on efficiency.

Most importantly, however, we reconcile quantum mechanics with local reality by modifying the handling of separated systems. We do not use the joint probability formula for cases of separated measurements; instead we use the marginals (partial traces or reduced density matrices) together with whatever priors we have from an understanding of the system. Specification of what are separated measurements is a somewhat delicate matter but this has a satisfactory answer that I develop elsewhere. If we accept this modification to quantum mechanics, nonlocality is eliminated. The experiments when correctly interpreted confirm the local realist position. The rest of quantum mechanics remains intact, and remains highly valued as a powerful probability calculus for observables. Without nonlocality to contend with, we can recruit powerful classical ideas, such as semiclassical radiation theory, stochastic dynamics, classical noncommutativity/contextuality, measurement effects on state, etc., to augment or complement quantum mechanics. The modified quantum mechanics can live in peaceful harmony with the local realist conception.

***Acknowledgement***: *I thank Gregor Weihs for making available the raw data files from the experiment, and for valuable discussion of the ideas contained here. I also thank Hans De Raedt for encouragement and helpful discussion.*